\documentclass[aps,prl,preprint,groupedaddress]{revtex4}

\usepackage{latexsym}           
\usepackage{amsmath}
\usepackage{amssymb}
\usepackage{color}

\newcommand{\etal}{{\it et al.}}

\newcommand{\stw}{\ensuremath{\sin ^{2} \theta_{\mathrm{w}}}}

\newcommand{\Qw}{\ensuremath{Q_{\mathrm{w}}}}

\newcommand{\qp}{\ensuremath{q_{\mathrm{p}}}}

\newcommand{\cs}{\ensuremath{^{133}\mathrm{Cs}}}

\newcommand{\z}{\ensuremath{Z^{0}}}

\renewcommand{\epsilon}{\ensuremath{\varepsilon}}

\newcommand{\psibar}{\ensuremath{\overline{\psi}}}

\newcommand{\me}{\ensuremath{m_{\mathrm{e}}}}

\newcommand{\baplus}{\ensuremath{\mathrm{Ba}^{+}}}

\newcommand{\hpnc}{\ensuremath{H_{\mathrm{PNC, 1}}}}

\newcommand{\Rpd}{\ensuremath{\mathcal{R}_{\mathrm{p-d}}}}

\begin{document}


\title{Estimate of Contribution from $P-D$ Mixing in Atomic PNC}


\author{M.~C.~Welliver}
\email[]{Marc.Welliver@Colorado.EDU}
\author{S.~J.~Pollock}
\email[]{Steven.Pollock@Colorado.EDU}
\affiliation{Dep't of Physics, University of Colorado, Boulder CO 80309}


\date{\today}


\begin{abstract}
  
  We investigate possible contributions to future atomic parity
  nonconservation (PNC) experiments from parity admixtures between
  single-electron atomic states with total angular momentum $j = 3/2$.
  We develop new formalism for studying these admixtures between
  atomic $p$ and $d$ states, which enter only when finite nuclear size
  effects are considered and have been neglected in the literature to
  date.  We use analytic approximations to provide an
  order-of-magnitude estimate of the contribution from these
  admixtures, and identify a dimensionless ratio which sets the scale
  of the correction.  Using realistic numerical wavefunctions in
  \baplus\ we confirm the results of our analytic expressions, and
  conclude quantitatively that these novel admixtures are likely to be
  negligible in essentially all cases.

\end{abstract}

\pacs{}

\maketitle



\section{Introduction}
\label{intro}

In recent years, precision experiments~\cite{wieman,fortson_tl}
measuring parity nonconservation (PNC) in atomic transitions have
developed into low-energy tests of the Standard Model which complement
high energy collider experiments.  PNC effects in atomic transitions
arise due to neutral-weak interactions between electrons and the
nucleus.  Atomic PNC experiments are sensitive to the radiatively
corrected weak charge of the nucleus, \Qw, which has been calculated
to one-loop in the Standard Model~\cite{marciano,rosner,langacker} and
is given at tree level by $\Qw^{0} = - N + Z ( 1 - 4 \stw )$.  Because
radiative corrections enter into low- and high-energy observables
differently, these precision atomic PNC experiments provide a search
for possible physics beyond the Standard Model which is complementary
to more conventional accelerator based searches.

It is well known~\cite{bouchiat,johnson,dzuba} that the experimental
observable must be corrected for finite nuclear size when compared to
the Standard Model prediction due to variation of the atomic electron
wavefunctions over the extent of the nucleus.  It has been
shown~\cite{fortson,pollock,npl-1163} that the transition amplitude
between initial and final atomic states can be formally factorized to
separate nuclear and atomic physics contributions.  The nuclear
structure correction to atomic PNC is non-negligible (about $4\%$ in
\cs) and grows as $(Z \alpha)^{2}$, but is reliably calculable.
Differences between neutron and proton distributions modify this
slightly~\cite{fortson,pollock,npl-1163}.

The most precise atomic PNC experiment to date~\cite{wieman} involves
parity admixtures of initial and final states with total angular
momentum $j = 1/2$, but there are several other PNC measurements which
have either been performed~\cite{fortson_tl} or are currently in
progress~\cite{fortson_ba,demille} looking for PNC effects in
transitions involving $j = 3/2$ final states.  To date, all
calculations in the
literature~\cite{bouchiat,fortson_ba,demille,fortson_tl} consider only
contributions from parity admixtures between $S_{1/2}$ and $P_{1/2}$
states because the $D_{3/2}$ state has vanishing amplitude and
derivative at the origin.  In principle, there could be an additive
contribution to the PNC transition amplitude arising from opposite
parity admixtures to the $j = 3/2$ final state which is induced by a
finite sized nucleus.  Motivated by the scale of the finite nuclear
size correction to $s-p$ mixing amplitudes, we have quantified the
effect of $p-d$ mixing when finite nuclear size effects are taken into
account.  

We begin this work by developing formalism necessary for studying
parity admixtures between $P_{3/2}$ and $D_{3/2}$ states in future
atomic PNC experiments.  We define a dimensionless ratio which sets
the scale of the additive contribution due to $p-d$ mixing relative to
the dominant $s-p$ mixing term.  We then use approximate analytic
calculations to provide an order-of-magnitude estimate of this ratio.
Using realistic numerical wavefunctions, we confirm our analytic
estimates in \baplus, one of the atomic systems in which PNC effects
involving a final $D$-state may be measured~\cite{fortson_ba}.

\section{Formalism}
\label{formalism}

The parity violating electron-nucleon interaction is dominated by
exchange of \z-bosons and can be written in terms of vector and
axial-vector currents for electrons and nucleons:
\begin{equation}
  \label{pnc_int}
  H_{\mathrm{PNC}} = A_{e} V_{\mathrm{N}} + V_{e} A_{\mathrm{N}} .
\end{equation}
The second term depends on the orientation of the nuclear spin and
usually amounts to at most a few percent of the first, in part because
the vector electron-\z\ coupling is small compared its axial-vector
counterpart.  Additional suppression arises because the nuclear
spin-dependent currents do not add coherently like the vector nucleon
currents, which are independent of the spin of the nucleus.  The
nuclear spin independent (NSID) piece of the interaction is
proportional to \Qw\ and can be isolated experimentally by averaging
over hyperfine states in the transition, so henceforth we neglect the
second term in Eq.~\ref{pnc_int}.

Due to the small effective momentum transfer associated with atomic
PNC observables, the three-vector part of the nuclear current in the
first term of Eq.~\ref{pnc_int} is highly suppressed relative to the
charge ($\mu = 0$) component.  Keeping only the dominant charge
component, we have~\cite{fortson,pollock,npl-1163}
\begin{equation}
  \label{pnc_amp}
  \begin{split}
    \langle i | \hpnc | j \rangle = & \frac{G_{\mathrm{F}}} {2
      \sqrt{2}} \int d^{3} r \psi^{\dagger}_{e \, i} ( {\mathbf r} ) 
    \gamma^{5} \psi_{e \, j} ( {\mathbf r} ) \rho_{\mathrm{w}} (
    {\mathbf r} ) \\ 
    = & \frac{G_{\mathrm{F}}} {2 \sqrt{2}} \mathcal{C}_{ij} ( Z )
    \mathcal{N} \qp\, \Qw^{\mathrm{exp}} .
  \end{split}
\end{equation}
Here $G_{\mathrm{F}}$ is the Fermi constant, $\rho_{\mathrm{w}}$ is
the weak nuclear charge density, $\mathcal{C}_{ij} ( Z )$ depends on
the full multi-electron wavefunction, and contains atomic structure
effects including many-body correlations for a point nucleus
calculation; $\mathcal{N} \equiv \psibar_{e \, i} (0) \gamma^{5}
\psi_{e \, j} (0)$ is a normalization factor for the single-electron
axial transition matrix element evaluated at the origin; \qp\ is the
correction factor for finite nuclear size.  The remaining factor,
$\Qw^{\mathrm{exp}}$, is the experimentally determined weak neutral
charge of the nucleus which is compared directly with the Standard
Model prediction.  The factorization in the second line of
Eq.~\ref{pnc_amp} can be used because the PNC matrix elements only
depend on the electronic wavefunctions over the extent of the nucleus
where binding energies can be neglected with respect to the Coulomb
potential.  Therefore, apart from an overall normalization factor, all
of the axial electronic matrix elements which contribute are
essentially identical.

Of the atomic PNC experiments involving $j = 3/2$ final states, the
proposed measurement~\cite{fortson_ba} of PNC in a single trapped
\baplus\ ion is the theoretically simplest, since it involves a single
valence electron outside of a tightly closed shell.  The others
involve at least two valence electrons, leading to additional
considerations such as configuration mixing which must ultimately be
taken into account.  In the proposed \baplus\ experiment, atomic PNC
effects will be measured by studying the allowed $\hat{E2}$ transition
between the $6 S_{1/2}$ ground state and the $5 D_{3/2}$ excited
state.  This transition has a small additional PNC-induced $\hat{E1}$
amplitude given by~\cite{fortson_ba}:
\begin{equation} \label{e_pnc_ba}
  \begin{split}
    \mathcal{E}_{m^{\prime} m}^{\mathrm{PNC}} & = \langle \overline{
      \rule{0mm}{3mm} 5 D_{3/2}, m^{\prime} } | \hat{E1} |
    \overline{ \rule{0mm}{3mm} 6 S_{1/2}, m } \rangle \\ 
    & \simeq \sum_{n, m^{\prime \prime}} \frac{ \langle 5 D_{3/2},
      m^{\prime} | \hat{E1} | n P_{1/2}, m^{\prime \prime} \rangle
      \langle n P_{1/2}, m^{\prime \prime} | \hpnc | 6 S_{1/2}, m
      \rangle }{ W_{6 S_{1/2}} - W_{n P_{1/2}} } \\ 
    & \quad + \sum_{n, m^{\prime \prime}} \frac{ \langle 5 D_{3/2}, 
      m^{\prime} | \hpnc | n P_{3/2}, m^{\prime \prime} \rangle
      \langle n P_{3/2}, m^{\prime \prime} | \hat{E1} | 6 S_{1/2}, m
      \rangle }{ W_{5D_{3/2}} - W_{nP_{3/2}} } \\ 
    & \equiv \mathcal{E}^{\mathrm{PNC}}_{\mathrm{s-p}} +
    \mathcal{E}^{\mathrm{PNC}}_{\mathrm{p-d}} ,
  \end{split}
\end{equation}
where $\mathcal{E}^{\mathrm{PNC}}_{\mathrm{s-p}}$ and
$\mathcal{E}^{\mathrm{PNC}}_{\mathrm{p-d}}$ carry implicit dependence
on $m^{\prime}$, $m$.  The $\Delta j = 0$ selection rule for matrix
elements of \hpnc\ has already been explicitly used in determining the
intermediate states present in the sums in this
expression~\cite{fortson,pollock,npl-1163}.  Here we have separated
the PNC-induced transition amplitude into contributions arising from
parity admixtures of the initial $s$-state
($\mathcal{E}^{\mathrm{PNC}}_{\mathrm{s-p}}$) and the final $d$-state
($\mathcal{E}^{\mathrm{PNC}}_{\mathrm{p-d}}$).

In initial estimates of the size of the PNC-induced amplitude for the
\baplus\ and similar experiments~\cite{fortson_ba,demille,fortson_tl}
only amplitudes arising from $s-p$ mixing were included, with the
assumption that $p$ and $d$ states are not mixed by \hpnc.  For a
point-like nucleus this is an exact description of the system (to
first order in \hpnc).  When finite nuclear size effects are included,
however, the validity of neglecting contributions from $p-d$ mixing is
less clear.  As discussed in
Refs.~\cite{fortson,pollock,npl-1163,welliver}, the neutral weak
interaction is a contact interaction between nucleons and the
transition electron in atomic PNC.  Therefore the matrix elements of
\hpnc\ depend on electronic wavefunctions evaluated over the extent of
the nucleus.  Upper- and lower-component Dirac radial wavefunctions
vanish as $r \rightarrow 0$ for $j \geq 3/2$, so all PNC matrix
elements in the second sum in Eq.~\ref{e_pnc_ba} vanish identically
for a point nucleus.  For a finite-sized nucleus the first term in
Eq.~\ref{e_pnc_ba} still dominates, but it has already been shown in
Refs.~\cite{pollock,npl-1163} that finite nuclear size effects can
modify $\mathcal{E}^{\mathrm{PNC}}_{\mathrm{s-p}}$ at the
$5-10\%$-level, depending on Z.  We now investigate the conditions
under which $\mathcal{E}^{\mathrm{PNC}}_{\mathrm{p-d}}$ might
contribute at levels approaching the finite nuclear size correction to
$\mathcal{E}^{\mathrm{PNC}}_{\mathrm{s-p}}$ in Eq.~\ref{e_pnc_ba}.

Since $\mathcal{E}^{\mathrm{PNC}}_{\mathrm{s-p}}$ is the dominant
piece it must be calculated quite accurately, including overall
normalizations which require sophisticated atomic many-body
calculations~\cite{johnson,dzuba}.  In order to study the contribution
from $p - d$ mixing, we factor out this dominant term, leaving a small
dimensionless correction factor which we estimate roughly:
\begin{equation} \label{p_d_factor}
  \mathcal{E}_{m' m}^{PNC} = \mathcal{E}^{\mathrm{PNC}}_{\mathrm{s-p}}
  \ \left[ 1 + \Rpd \right] , 
\end{equation}
where
\begin{equation} \label{p_d_ratio}
  \Rpd \equiv \frac{ \mathcal{E}^{\mathrm{PNC}}_{\mathrm{p-d}} }{ 
    \mathcal{E}^{\mathrm{PNC}}_{\mathrm{s-p}} } . 
\end{equation}
Formally, this ratio involves sums over transition matrix elements of
\hpnc\ and $\hat{E1}$ operators between full many-electron atomic
states including correlation effects, the calculation of which is
beyond the scope of the present work.  Since our goal is an
order-of-magnitude estimate for the correction, however, we
approximate the transition matrix elements using single-electron
atomic states of nominally good parity.  We expect this to be a
reasonable approximation for \baplus\ because the transition involves
a single valence electron outside of a tightly bound core.  Using the
solutions to the Dirac central potential problem with definite
parity~\cite{bjorken_drell}, we perform the angular integration
explicitly~\cite{welliver}, reducing these single-electron matrix
elements to expressions containing simple radial integrals.  We then
express the PNC matrix elements as
\begin{align} \label{s_p_me}
  \langle n P_{1/2}, m^{\prime \prime} | H_{\mathrm{PNC}} | 6 S_{1/2},
  m \rangle & \equiv - i \ \delta_{m^{\prime \prime} m} \frac{ G_{F}
  }{ 2 \sqrt{2} } \ \mathcal{I}^{\mathrm{PNC}}_{np 6s} \\ 
  \label{p_d_me}
  \langle 5 D_{3/2}, m^{\prime} | H_{\mathrm{PNC}} | n P_{3/2},
  m^{\prime \prime} \rangle & \equiv - i \ \delta_{m^{\prime}
    m^{\prime \prime}} \frac{ G_{F} }{ 2 \sqrt{2} } \
  \mathcal{I}^{\mathrm{PNC}}_{5d np} , 
\end{align}
and the $\hat{E1}$ matrix elements as (taking the $\hat{z}$-component
for simplicity)
\begin{align} \label{e1_p_d_me}
  \langle 5 D_{3/2}, m^{\prime} | \hat{E1}_{z} | n P_{1/2}, m^{\prime
    \prime} \rangle & \equiv i \frac{ 2 \sqrt{2} }{ 3 } \ e \ 
  \delta_{m^{\prime} m^{\prime \prime}} \ \mathcal{I}^{E1}_{5d np} \\ 
  \label{e1_s_p_me}
  \langle n P_{3/2}, m^{\prime \prime} | \hat{E1}_{z} | 6 S_{1/2}, m
  \rangle & \equiv i \frac{ 2 \sqrt{2} }{ 3 } \ e \ \delta_{m^{\prime 
      \prime} m} \ \mathcal{I}^{E1}_{np 6s} \ .
\end{align}
With these simplifications applied to Eqs.~\ref{e_pnc_ba}
and~\ref{p_d_factor}, our correction factor arising from $p-d$ mixing
can be written
\begin{equation} \label{p_d_ratio_simp_1}
  \Rpd = \frac{ \sum_{n} \mathcal{I}^{\mathrm{PNC}}_{5dnp}
    \mathcal{I}^{E1}_{np6s} / (W_{5 D_{3/2}} - W_{n P_{3/2}}) }{
    \rule{0mm}{4mm} \sum_{n} \mathcal{I}^{E1}_{5dnp}
    \mathcal{I}^{\mathrm{PNC}}_{np6s} / (W_{6 S_{1/2}} - W_{n
      P_{1/2}}) } . 
\end{equation}
In these expressions we find it convenient to define transition radial
integrals, $\mathcal{I}_{n_{1} l_{1} n_{2} l_{2}}$, because phases,
numerical constants, and all dependence on magnetic quantum numbers
cancel in the ratio.

We now turn to evaluating the transition radial integrals of interest.
In terms of single-electron radial wavefunctions, the integrals
arising from the matrix elements of \hpnc\ take the form
  \begin{gather} \label{s_p_pnc_int}
    \mathcal{I}^{\mathrm{PNC}}_{np 6s} = \int d r \rho_{\mathrm{w}} 
    (r) \left[ G_{n p_{1/2}} (r) F_{6 s_{1/2}} (r) - F_{n p_{1/2}} (r)
      G_{6 s_{1/2}} (r) \right] \\ 
    \label{p_d_pnc_int}
    \mathcal{I}^{\mathrm{PNC}}_{5d np} = \int d r \rho_{\mathrm{w}} 
    (r) \left[ G_{5 d_{3/2}} (r) F_{n p_{3/2}} (r) - F_{5 d_{3/2}} (r)
      G_{n p_{3/2}} (r) \right] . 
  \end{gather}
These integrals constitute the finite nuclear size corrections to the
$s-p$ and $p-d$ mixing contributions to atomic PNC, with the weak
charge density of the nucleus ($\rho_{\mathrm{w}} = - N
\rho_{\mathrm{n}} + Z (1 - 4 \stw) \rho_{\mathrm{p}}$) integrated
against a radial folding function.  These expressions contain
essentially all dependence of \Rpd\ on nuclear physics.  The
transition radial integrals arising from the $\hat{E1}$ matrix
elements are given by~\cite{welliver}
\begin{align} \label{p_d_e1_int}
  \mathcal{I}^{E1}_{5d np} & = \int d r \ F_{5 d_{3/2}} (r) G_{n
    p_{1/2}} (r) \\
  \label{s_p_e1_int}
  \mathcal{I}^{E1}_{np 6s} & = - \int d r \ G_{n p_{3/2}} (r)
  F_{6 s_{1/2}} (r) .
\end{align}
The parity-allowed $\hat{E1}$ transition strengths and energy
denominators present in the ratio, \Rpd, are generally of similar
orders of magnitude~\cite{baik,johnson_wfns}.  Therefore we expect the
scale of the correction due to $p-d$ mixing to be set by the PNC
transition matrix elements.  We now look for analytic approximations
to the transition radial integrals in Eqs.~\ref{s_p_pnc_int}
and~\ref{p_d_pnc_int}.

\section{Approximate Analytic Results}
\label{approx_res}

Although the integrals in Eqs.~\ref{s_p_pnc_int} and~\ref{p_d_pnc_int}
formally extend over all space, the weak charge density only
contributes appreciably for $r \lesssim 10~\mathrm{fm}$.  Therefore
the PNC transition radial integrals can be estimated reliably by
considering the form of the electron wavefunctions in the vicinity of
the nucleus.  In this region the electronic potential is dominated by
the nuclear Coulomb potential, with screening and correlation effects
negligible~\cite{fortson}.  Also, the electronic binding energies can
be safely neglected with respect to the potential in the vicinity of
the nucleus, as discussed in Refs.~\cite{npl-1163,welliver}.  Using
power series solutions of the Dirac radial
equations~\cite{bjorken_drell} under these minimal assumptions, we
find that the PNC transition radial integrals can be expressed as
\begin{gather} \label{s_p_pnc_expand}
  \begin{split}
    \mathcal{I}^{\mathrm{PNC}}_{np 6s} \simeq \frac{1}{4 \pi} A_{n
      p_{1/2}} A_{6 s_{1/2}} & \left\{ - N \left[ 1 - \frac{2}{9} 
        \phi_{0} (\phi_{0} - 2 \me) \langle r^{2}
        \rangle_{\mathrm{n}} + \cdots \right] \right. \\ 
    & \left. \ \, + Z (1 - 4 \stw) \left[ 1 - \frac{2}{9} \phi_{0} 
        (\phi_{0} - 2 \me) \langle r^{2} \rangle_{\mathrm{p}} +
        \cdots \right] \right\} 
  \end{split} \\
  \label{p_d_pnc_expand}
  \mathcal{I}^{\mathrm{PNC}}_{5d np} \simeq \frac{ 1 }{ 4 \pi }
  A_{5 d_{3/2}} A_{n p_{3/2}} \left\{ - N \langle r^{2}
    \rangle_{\mathrm{n}} + Z (1 - 4 \stw) \langle r^{2} 
    \rangle_{\mathrm{p}} + \cdots \right\} . 
\end{gather}
In obtaining these results we have expanded the realistic nuclear
Coulomb potential in a Taylor series about the origin, $V(r) \simeq
\phi_{0} + \phi_{2} r^{2} + \cdots$.  Here $\phi_{0}$ is the leading
term in the expansion of the potential, and $A_{nlj}$ is the
coefficient of the leading term in a power series expansion of the
normalized single-electron wavefunction, $|nlj, m \rangle$.  We note
that the quantities in square brackets in Eq.~\ref{s_p_pnc_expand} are
the finite nuclear size corrections defined in
Refs.~\cite{fortson,pollock,npl-1163}, and that the $p-d$ transition
radial integrals, $\mathcal{I}^{\mathrm{PNC}}_{5d np}$, vanish for a
point nucleus as expected.

Because we are looking for an order-of-magnitude estimate of \Rpd,
details like the \mbox{$\sim 5\%$} finite nuclear size correction to
$\mathcal{I}^{\mathrm{PNC}}_{np 6s}$ and possible differences in
neutron and proton mean-square radii are unimportant at this stage.
We further simplify the PNC transition radial integrals by setting the
finite nuclear size correction equal to unity in
Eq.~\ref{s_p_pnc_expand} and replacing $\langle r^{2}
\rangle_{\mathrm{n}}$ and $\langle r^{2} \rangle_{\mathrm{p}}$ in
Eq.~\ref{p_d_pnc_expand} with the nucleon mean-square radius $\langle
r^{2} \rangle_{\mathrm{N}}$.  With these approximations we find
\begin{equation} \label{pnc_ints_approx}
  \begin{split}
    \mathcal{I}^{\mathrm{PNC}}_{np 6s} & \approx \frac{ 1 }{ 4 \pi }
    A_{n p_{1/2}} A_{6 s_{1/2}} \left\{ -N + Z (1 - 4 \stw) \right\}
    \\ 
    \mathcal{I}^{\mathrm{PNC}}_{5d np} & \approx \frac{ 1 }{ 4 \pi }
    A_{5 d_{3/2}} A_{n p_{3/2}} \left\{ - N + Z (1 - 4 \stw) \right\}
    \langle r^{2} \rangle_{\mathrm{N}} . 
  \end{split}
\end{equation}
It is important to note that the $A_{nlj}$'s in
Eqs.~\ref{s_p_pnc_expand} and~\ref{p_d_pnc_expand} are dimensionful
constants, and that the constants for $j = 1/2$ and $j = 3/2$ have
different units.  Both $\mathcal{I}^{\mathrm{PNC}}_{np 6s}$ and
$\mathcal{I}^{\mathrm{PNC}}_{5d np}$ arise from matrix elements of the
same operator, \hpnc, so the ratio must be dimensionless.  This
expansion of the integrals for small $r$ shows, therefore, that the
ratio $(A_{5 d_{3/2}} A_{n_{1} p_{3/2}}) / (A_{n_{2} p_{1/2}} A_{6
  s_{1/2}})$ should have dimension $[L]^{-2}$, where $L$ is a length
scale relevant to the normalization of atomic bound state
wavefunctions.

This dependence on a characteristic atomic length scale is indicative
of the fundamental difference between estimating contributions from
$p-d$ mixing and studying the sensitivity of the Boulder \cs\ 
experiment to the spatial neutron
distribution~\cite{npl-1163,welliver}.  In the Boulder
experiment~\cite{wieman}, only parity admixtures between $s$ and $p$
states are present, so a formal factorization of the transition
amplitude (Eq.~\ref{pnc_amp}) is possible.  This factorization gives
separate multiplicative factors dependent on contributions from
nuclear and atomic physics which can be calculated independently.  In
evaluating \Rpd\ no such factorization is possible, so we must
estimate the normalization constants $A_{nlj}$.  Plugging
Eqs.~\ref{pnc_ints_approx} into our expression for the $p-d$ mixing
correction factor, Eq.~\ref{p_d_ratio_simp_1}, we now have
\begin{equation}
  \label{p_d_ratio_simp_2}
  \Rpd \approx \frac{ A_{5 d_{3/2}} \langle r^{2} \rangle_{\mathrm{N}}
    }{ A_{6 s_{1/2}} } \frac{ \sum_{n} A_{n p_{3/2}}
      \mathcal{I}^{E1}_{np6s} / (W_{5 D_{3/2}} - W_{n P_{3/2}}) }{ 
      \rule{0mm}{4mm} \sum_{n} A_{n p_{1/2}} \mathcal{I}^{E1}_{5dnp} /
      (W_{6 S_{1/2}} - W_{n P_{1/2}}) } . 
\end{equation}
Here, the sums in numerator and denominator should be of similar
orders of magnitude~\cite{baik,johnson_wfns}, so we expect that the
overall scale of the correction factor will be given by the ratio out
in front.  We will confirm this expectation numerically for \baplus\ 
in the next section.

In order to estimate the coefficients $A_{nlj}$ in
Eqs.~\ref{s_p_pnc_expand} and~\ref{p_d_pnc_expand}, normalized
electronic wavefunctions satisfying the condition $\int_{0}^{\infty}
\left( G_{nlj}^{2} + F_{nlj}^{2} \right) = 1$ must be constructed and
expanded for small $r$.  Hence, calculation of the normalization
coefficients at the origin requires knowledge of the electron
wavefunctions over atomic distance scales where shielding and
correlation effects become important.  Such calculations are beyond
the scope of the present work.  We note, however, that our correction
factor \Rpd\ will depend on the ratio of $\langle r^{2}
\rangle_{\mathrm{N}}$ to the relevant atomic length scale, and that
all atoms are roughly the same size.  We therefore first look to
hydrogenic electron wavefunctions for analytic expressions to estimate
the $A_{nlj}$ and study the dependence on the characteristic
parameters of the problem.  Fractional deviations between
(upper-component)ß relativistic and nonrelativistic radial
wavefunctions are generally of order $(Z \alpha)^{2}$, except where
one of the wavefunctions has a node~\cite{rose}.  We therefore look at
the normalization of nonrelativistic hydrogenic wavefunctions at the
origin for order-of-magnitude estimates of the $A_{nlj}$'s.  The
coefficient of the leading term in an expansion of the nonrelativistic
hydrogenic wavefunction as $r \rightarrow 0$ is given
by~\cite{welliver,non_rel_texts}
\begin{equation}
  \label{non_rel_coeff}
  C_{nl} = \left( \frac{ 2 Z }{ n a_{0} } \right)^{3/2 + l} \left[
    \frac{ (n + l)! }{ 2 n (n - l - 1)! } \right]^{1/2} \frac{ 1 }{ (
    2 l + 1 )! } ,
\end{equation}
where $a_{0}$, the Bohr radius, gives the length scale relevant to
atomic wavefunction normalizations.  Replacing the $A_{nlj}$'s in
Eq.~\ref{p_d_ratio_simp_2} by the nonrelativistic $C_{n l}$'s we find
\begin{equation}
  \label{p_d_ratio_simp_3}
  \begin{split}
    \Rpd & \simeq 0.04 \ \frac{ Z^{2} \langle r^{2}
      \rangle_{\mathrm{N}} }{ a_{0}^{2} } \ \frac{ \sum_{n} C_{n
        p} \mathcal{I}^{E1}_{np6s} / (W_{5 D_{3/2}} - W_{n P_{3/2}})
    }{ \rule{0mm}{4mm} \sum_{n} C_{n p} \mathcal{I}^{E1}_{5dnp} /
      (W_{6 S_{1/2}} - W_{n P_{1/2}}) } \\ 
    & \sim 1 \times 10^{-6} \quad \frac{ \sum_{n} C_{n p}
      \mathcal{I}^{E1}_{np6s} / (W_{5 D_{3/2}} - W_{n P_{3/2}}) }{ 
      \rule{0mm}{4mm} \sum_{n} C_{n p} \mathcal{I}^{E1}_{5dnp} /
      (W_{6 S_{1/2}} - W_{n P_{1/2}}) } ,
  \end{split}
\end{equation}
(using $Z = 56$ and $\langle r^{2} \rangle_{\mathrm{N}}^{1/2} =
4.84~\mathrm{fm}$ for \baplus).  Based on this rough estimate, it
appears that the ratio of nuclear to atomic length scales predicted by
the analytic approximations serves as the primary suppression factor
in considering $p-d$ mixing.  Thus far we have made general arguments
that the scale of \Rpd\ should be set by the ratio of PNC radial
integrals in Eq.~\ref{p_d_ratio_simp_1}, without estimating the
$\hat{E1}$ radial integrals or energy denominators.  Physically, we
know that $\hat{E1}$ transition strengths and binding energies for the
states of interest here are generally within an order of magnitude of
one another~\cite{baik,johnson_wfns}.  Theoretically, these quantities
cannot be reliably estimated in the context of our approximations
because they depend on long range atomic physics.  In the next section
we use realistic numerical wavefunctions~\cite{johnson_wfns} to
evaluate all quantities in Eq.~\ref{p_d_ratio_simp_1} as a check of
our approximate analytic results.

\section{Numerical Results}
\label{baplus_wfns}

The calculations presented in the previous section are instructive in
quantifying the mechanism of suppression for parity admixtures of
atomic states with $j \geq 3/2$, but are admittedly crude.  We now
wish to test the validity of the order-of-magnitude estimate as well
as investigate whether the energy denominators and $\hat{E1}$
transition radial integrals which we have not yet addressed might give
rise to enhancements to \Rpd.  We have obtained realistic binding
energies and atomic radial wavefunctions for \baplus, calculated in a
relativistic Dirac-Hartree-Fock model~\cite{johnson_wfns}.  Using
these tabulated wavefunctions, we have performed the transition radial
integrals in Eqs.~\ref{s_p_pnc_int} through~\ref{s_p_e1_int}
numerically.  The sums in both numerator and denominator of
Eq.~\ref{p_d_ratio_simp_1} are dominated by the $n = 6$ admixed
states, but we have included contributions from $n = 7,8$ as well.
This calculation yields $\Rpd \approx 4 \times 10^{-6}$, in good
agreement with our crude but analytic estimate.

In order to study this numerical check on our approximate analytic
calculation, we have fit the relativistic numerical wavefunctions to
polynomials near the origin and extracted the coefficients of the
leading terms of the upper- and lower-components.  Overall
wavefunction normalizations are not fixed by power series solutions
near the origin, but the ratio of leading term normalizations of
upper- and lower-component wavefunctions is uniquely determined by the
value of the potential at the origin.  Comparing this ratio estimated
from the fitted numerical wavefunctions and from power series
solutions for relativistic hydrogenic wavefunctions assuming that the
total potential at the origin is dominated by the nuclear Coulomb
potential, we find that the two methods agree to better than $5\%$.
As an additional check, we assume that \Rpd\ is dominated by the $n=6$
term and estimate $\mathcal{I}^{\mathrm{PNC}}_{6dnp} /
\mathcal{I}^{\mathrm{PNC}}_{6p6s}$ as simplified in
Eqs.~\ref{pnc_ints_approx} using both nonrelativistic and fitted
coefficients.  The nonrelativistic hydrogenic estimate gives $1 \cdot
10^{-6}$, while the estimate from fitting numerical wavefunctions
gives $0.6 \cdot 10^{-6}$.  These numerical checks indicate that our
crude analytic approximate results are in reasonable agreement with a
more detailed, full calculation requiring many-electron atomic
wavefunctions including the effects of correlations and shielding.

\section{Conclusions}
\label{conclusions}

These calculations indicate that neglecting possible $p-d$ mixing
contributions to atomic PNC amplitudes which involve $j = 3/2$ initial
or final states is an exceedingly good approximation.  We have
demonstrated that errors from neglecting $p-d$ mixing effects are well
below other sub-$1\%$ effects which have been neglected to date,
including contributions from the three-vector nucleon currents,
radiative corrections such as one-photon one-\z\ exchange box
diagrams, and parity violating electron-electron interactions.  While
we have focused on the simplest atomic system proposed for experiment,
\baplus, we expect this result to be applicable to more complex
systems like Ytterbium~\cite{demille} because the scale of the
correction is proportional to $Z^{2} \langle r^{2}
\rangle_{\mathrm{N}} / a_{0}^{2}$ in all cases.  The most likely
source of significant enhancement of the $p-d$ mixing term would be an
exceedingly small energy difference between the the admixed $p$ and
$d$ states in a heavy atom, which would depend sensitively on the
particular atomic system.



\begin{thebibliography}{999}


\bibitem{wieman} C.~S.~Wood, S.~C.~Bennett, D.~Cho, B.~P.~Masterson, 
  J.~L.~Roberts, C.~E.~Tanner, and C.~E.~Wieman,
  Science {\bf 275}, 1759 (1997); 
  S.~C.~Bennett and C.~E.~Wieman,
  Phys. Rev. Lett. {\bf 82}, 2484 (1999).

\bibitem{fortson_tl} P.~A.~Vetter, \etal,
  Phys. Rev. Lett. {\bf 74}, 2658 (1995); 
  N.~H.~Edwards, \etal,
  Phys. Rev. Lett. {\bf 74}, 2654 (1995).

\bibitem{marciano} W.~Marciano and J.~Rosner,
  Phys. Rev. Lett. {\bf 65}, 2963 (1990); {\bf 68}, 898(E) (1992).
  
\bibitem{rosner} J.~Rosner,
  Phys. Rev. {\bf D42}, 3107 (1990); {\bf D53}, 2724 (1996); 
  {\bf D61}, 016006 (1999).
  
\bibitem{langacker}  Particle Data Group,
  Euro. Phys. J. {\bf C54}, 1, (2000).

\bibitem{bouchiat} M.~A.~Bouchiat and C.~Bouchiat,
  Phys. Lett. {\bf B48}, 111 (1974); Rep. Prog. Phys. {\bf 60}, 1351 
  (1997). 
  
\bibitem{johnson} S.~A.~Blundell, J. Sapirstein, W.~R.~Johnson, 
  Phys. Rev. {\bf D45} 1602 (1992).

\bibitem{dzuba} V.~A.~Dzuba \etal, 
  Phys Lett. {\bf A 141}, 147 (1989).
  
\bibitem{fortson} E.~N.~Fortson, Y.~Pang, and L.~Wilets,
  Phys. Rev. Lett. {\bf 65}, 2857 (1990).

\bibitem{pollock} S.~J.~Pollock, E.~N.~Fortson, and L.~Wilets,
  Phys. Rev. {\bf C46}, 2587 (1992).
  
\bibitem{npl-1163} S.~J.~Pollock and M.~C.~Welliver,
  Phys. Lett. {\bf B464}, 177 (1999).

\bibitem{fortson_ba} N.~Fortson,
  Phys. Rev. Lett. {\bf 70}, 2383 (1993).

\bibitem{demille} D.~DeMille,
  Phys. Rev. Lett. {\bf 74}, 4165 (1995).

\bibitem{welliver} M.~Welliver,
  Ph.~D. thesis, University of Colorado, Boulder (2002).

\bibitem{bjorken_drell} J.~D.~Bjorken and S.~D.~Drell, 
  {\bf Relativistic Quantum Mechanics}, McGraw-Hill, New York (1964). 

\bibitem{baik} D.~H.~Baik \etal,
  At. Data Nucl. Data Tables {\bf 47}, 177 (1991).

\bibitem{johnson_wfns} W.~R.~Johnson, private communication.

\bibitem{rose} M.~E.~Rose,
  {\bf Relativistic Electron Theory}, John Wiley \& Sons, New York 
  (1961). 

\bibitem{non_rel_texts} S.~Gasiorowicz,
  {\bf Quantum Physics}, John Wiley \& Sons, New York (1996); 
  J.~J.~Sakurai,
  {\bf Modern Quantum Mechanics}, Addison-Wesley, Reading, MA (1994). 


\end{thebibliography}

\end{document}